\documentclass[aps,prl,twocolumn,epsfig]{revtex4}
\usepackage{latexsym}
\usepackage{graphicx}

\begin{document}

\title{Geometry of random interactions}
\author{P.~Chau Huu-Tai$^1$,
A.~Frank$^{2,3}$, N.~A.~Smirnova$^4$, and P.~Van Isacker$^1$}
\affiliation{$^1$Grand Acc\'el\'erateur National d'Ions Lourds,\\
Bo\^\i te Postale 55027, F-14076 Caen Cedex 5, France}
\affiliation{$^2$Instituto de Ciencias Nucleares,
Universidad Nacional Aut\'onoma de M\'exico,\\
Apartado Postal 70-543, 04510 M\'exico, D.F., M\'exico}
\affiliation{$^3$Centro de Ciencias F\'\i sicas,
Universidad Nacional Aut\'onoma de M\'exico,\\
Apartado Postal 139-B, 62251 Cuernavaca, Morelos, M\'exico}
\affiliation{$^4$University of Leuven, Instituut voor Kern- en Stralingsfysica,\\
Celestijnenlaan 200D, B-3001 Leuven, Belgium}

\begin{abstract}
It is argued that spectral features
of quantal systems with random interactions
can be given a geometric interpretation.
This conjecture is investigated
in the context of two simple models:
a system of randomly interacting $d$ bosons
and one of randomly interacting fermions
in a $j={7\over2}$ shell.
In both examples the probability for a given state
to become the ground state
is shown to be related to a geometric property
of a polygon or polyhedron
which is entirely determined by particle number, shell size,
and symmetry character of the states.
Extensions to more general situations are discussed.
\end{abstract}
\maketitle


\noindent
Recent studies
in the nuclear shell model~\cite{JON98,BIJ99,ZHA01,ZHA01b}
and the interacting boson model~\cite{BIJ00,KUS00}
with random interactions
have unveiled a high degree of order.
In particular, a marked statistical preference
was found for ground states with $J=0^+$.
In this Letter it is argued
that spectral features of quantal systems with random interactions
can be given a geometric interpretation
which allows the computation of the probability
for the quantum mechanical ground state
to have a specific angular momentum,
based on purely geometric considerations.
Although these results are obtained
in the context of a variety of simple models
which do not cover the full complexity of random interactions,
we believe them to be sufficiently general
to conjecture the possibility
of an entirely geometric analysis of the problem.

Consider a system
consisting of $n$ interacting particles (bosons or fermions)
carrying angular momentum $j$,
integer or half-integer.
Eigenstates of a rotationally invariant Hamiltonian
are characterized as $|j^n\alpha JM\rangle$
where $J$ and $M$ are the total angular momentum
and its projection,
and $\alpha$ is any other index
needed for a complete labeling of the state.
Although spectral properties
of a Hamiltonian with both one- and two-body interactions
can be analyzed in the way explained below,
we assume for simplicity
that the one-body contribution
is constant for all eigenenergies
and that the energy spectrum
is generated by two-body interactions only.
Under this assumption its matrix elements can be written as
\begin{equation}
\langle j^n\alpha J|\hat H|j^n\alpha'J\rangle=
{{n(n-1)}\over2}
\sum_Lc^L_{n\alpha J}c^L_{n\alpha'J}
G_L,
\label{matrix}
\end{equation}
where $M$ is omitted
since energies do not depend on it.
The quantities $G_L$ are two-particle matrix elements,
$G_L\equiv\langle j^2L|\hat H_2|j^2L\rangle$,
and completely specify the two-body interaction
while $c^L_{n\alpha J}$
are interaction-independent coefficients.
They can be expressed
in terms of coefficients of fractional parentage (CFP)~\cite{TAL93}
and, as such, are entirely determined
by the symmetry character of the $n$-particle states.

We begin by considering the special case
when a basis $|j^n\alpha JM\rangle$ can be found
in which the expansion~(\ref{matrix}) is diagonal.
In this case the Hamiltonian matrix elements
reduce to the energy eigenvalues
\begin{equation}
E_{n\alpha J}=
{{n(n-1)}\over2}
\sum_Lb^L_{n\alpha J}G_L,
\label{energ}
\end{equation}
with $b^L_{n\alpha J}\equiv\left(c^L_{n\alpha J}\right)^2$.
This is obviously a simplification
of the more general problem~(\ref{matrix})
but nevertheless a wide variety of simple model situations
can be accommodated by it.
For example, this property is valid
for any interaction between identical fermions
if $j\leq{7\over2}$
and remains so approximately for larger $j$ values;
it is also exactly valid for $p$, $d$, or $f$ bosons.
We shall refer to this class of problems as diagonal.
For a constant interaction, $G_L=1$,
all $n$-particle eigenstates
are degenerate with energy ${1\over2}n(n-1)$
and consequently the coefficients $b^L_{n\alpha J}$
satisfy the properties $\sum_Lb^L_{n\alpha J}=1$
and $0\leq b^L_{n\alpha J}\leq1$.
Equation~(\ref{energ}) can thus be rewritten
in terms of scaled energies as
\begin{equation}
e_{n\alpha J}\equiv
{{2E_{n\alpha J}}\over{n(n-1)}}=
G_{L'}+\sum_Lb^L_{n\alpha J}(G_L-G_{L'}),
\label{energy}
\end{equation}
for arbitrary $L'$.
This shows that,
in the case of $N$ interaction matrix elements $G_L$,
the energy of an arbitrary eigenstate
can, up to a constant scale and shift,
be represented as a point in a vector space
spanned by $m\equiv N-1$ differences of matrix elements.
Note that the position of these states
is fixed by $b^L_{n\alpha J}$ and hence interaction-independent.
Furthermore, since $0\leq b^L_{n\alpha J}\leq1$,
all states are confined to a compact region of this space
with the size of one unit in each direction.
For independent variables $G_L$ with covariance matrix
$\langle G_LG_{L'}\rangle=\delta_{LL'}$,
states are represented in an orthogonal basis.
The differences in~(\ref{energy}) are not independent
but have a covariance matrix of the form $1+\delta_{LL'}$;
this leads to a representation in a $m$-simplex basis
(i.e., an equilateral triangle in $m=2$ dimensions,
a regular tetrahedron for $m=3$,\dots).

The following procedure can now be proposed
to determine the probability $P_{n\alpha J}$
for a specific state $n\alpha J$ to become the ground state.
First, construct all points
corresponding to the energies $e_{n\alpha J}$.
Next, build from them the largest possible {\em convex} polytope
(i.e., convex polyhedron in $m$ dimensions).
All points (i.e.~states) inside this polytope
can {\em never} be the ground state
for {\em whatever} choice of matrix elements $G_L$
and thus have $P_{n\alpha J}=0$.
Finally, the probability $P_{n\alpha J}$ of any other state
at a vertex of the polytope
is a function of some geometric property at that vertex.

\begin{figure}[htbp]
\begin{center}
\includegraphics[height=7.2cm,width=6cm]{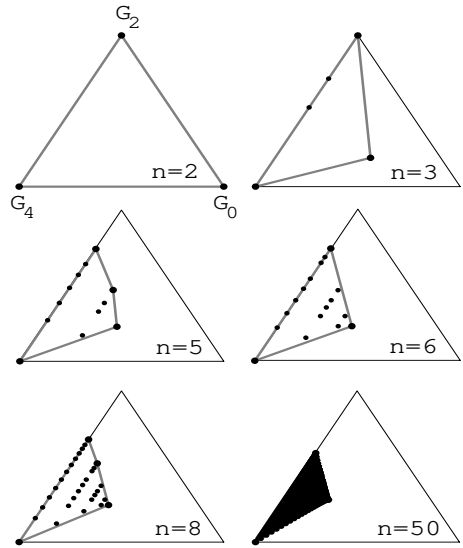}
\caption{Polygons corresponding to several systems of interacting $d$ bosons. All states are represented by a dot. The
smaller dots are inside or on the edge of a convex polygon (with the larger dots as vertices and indicated in grey) and are never the ground state. The probability of the vertex
states to be the ground state is related to the angle at the vertex.} \label{triangles}
\end{center}
\end{figure}

This general procedure can be illustrated with some examples.
Consider first a system of $j=2$ $d$ bosons.
In this case there are three interaction matrix elements, $G_0$, $G_2$,
and $G_4$ with $G_L\equiv\langle d^2L|\hat H_2|d^2L\rangle$;
thus, $N=3$ and the problem can be represented in a plane.
Because of the ${\rm U}(5)\supset{\rm SO}(5)\supset{\rm SO}(3)$
algebraic structure,
an analytic solution is known for $n$ interacting $d$ bosons
with eigenenergies~\cite{BRI65,ARI76,CAS78}
\begin{eqnarray}
e_{n\tau J}&=&
{{2n(n-2)+2\nu(\nu+3)-J(J+1)}\over{7n(n-1)}}(G_2-G_4)
\nonumber\\
&+& {{n(n+3)-\nu(\nu+3)}\over{5n(n-1)}} (G_0-G_4)+G_4,
\label{denerg}
\end{eqnarray}
where $\nu$ is the seniority quantum number
which counts the number of $d$ bosons not in pairs
coupled to angular momentum zero.
Any energy $e_{n\nu J}$ can be represented as a point
inside an equilateral triangle with vertices $G_0$, $G_2$, and $G_4$
of which the position is determined by the appropriate values
on the edges $G_0-G_4$ and $G_2-G_4$.
Examples for several boson numbers $n$
are shown in Fig.~\ref{triangles}.
For $n=2$ there are three states
with $L=0,2,4$ and energies $e=G_0,G_2,G_4$;
clearly, they have equal probability of being the ground state.
As $n$ increases, more states appear in the triangle.
The majority of states, shown as the smaller dots,
are inside the convex polygon indicated in grey
and can never be the ground state
for whatever the choice of $G_L$.
If we translate or rotate the convex polygon
inside the triangle,
its properties should not change
since the points $G_0$, $G_2$, and $G_4$ are equivalent and since the distribution depends only on $\sum_i G_i^2$.
Thus the probability for a point to be the ground state
can only be related to the angle
subtended at that vertex.
The relation can be formally derived
but also inferred from a few simple examples.
If the polygon is an equilateral triangle, square, regular pentagon,\dots
each vertex is equally probable
with probability $1\over3$, $1\over4$, $1\over5$,\dots
One deduces the relation
(see the $m=2$ polygons in Table~\ref{polytopes})
\begin{equation}
P^{(2)}_v={1\over2}-{{\theta_v}\over{2\pi}}
\label{p2d}
\end{equation}
between the angle $\theta_v$ at the vertex $v$ of the polygon
and the probability $P^{(2)}_v$ for the state associated
with that vertex to be the ground state.
\begin{table}
\centering \caption{The angle or the angle sum $\sum_{f\ni v}\theta_{vf}$ and
 the vertex probability $P_v$ of regular polygons ($m=2$) and polyhedrons ($m=3$).} \label{polytopes}
\smallskip
\begin{tabular}{rrcc|rrcc}
\hline 
\hline 
  & $m$&$\displaystyle{\sum_{f\ni v}\theta_{vf}}$& $P_v$ &  & $m$&
$\displaystyle{\sum_{f\ni v}\theta_{vf}}$& $P_v$\\
\hline
triangle    & 2&     ${1\over3}\pi$&     $1\over3$ & tetrahedron  &3&              $\pi$&     $1\over4$\\
square      & 2&     ${1\over2}\pi$&     $1\over4$ & octahedron  & 3&     ${4\over3}\pi$&     $1\over6$\\
pentagon    & 2&     ${3\over5}\pi$&     $1\over5$ & icosahedron & 3&     ${5\over3}\pi$&  $1\over{12}$\\
hexagon    & 2&     ${2\over3}\pi$&     $1\over6$ & cube        & 3&     ${3\over2}\pi$&     $1\over8$\\
$p$-gon     & 2&${{p-2}\over p}\pi$&    $1\over p$ & dodecahedron& 3&     ${9\over5}\pi$&  $1\over{20}$\\
\hline 
\hline 
\end{tabular}
\end{table}
Table~\ref{dboson} compares
probabilities calculated with the analytic relation~(\ref{p2d})
with those obtained from numerical tests
for several systems of randomly interacting $d$ bosons
with boson numbers $n=5,6,10,18$.
The numerical probabilities are obtained
from 20000 runs with random interaction parameters.
They agree with the analytic result~(\ref{p2d}).
\begin{table}
\centering
\caption{Probabilities $P_{n\nu J}$ (in \%) of some $d$-boson states
obtained analytically with formula~(\protect\ref{p2d})
and numerically with Gaussian parameters.}
\label{dboson}
\smallskip
\begin{tabular}{ccrr|ccrr}
\hline
\hline
&&\multicolumn{2}{c|}{Probability $P_{n\nu J}$} &  & &\multicolumn{2}{c}{Probability $P_{n\nu J}$} \\
\cline{3-4}\cline{7-8}
$n$&$J(\nu)$&Analytic&Gaussian & 
$n$&$J(\nu)$&Analytic&Gaussian \\
\hline 
 5&   0(3)&  4.08 &  3.96 &10&   0(0)& 20.87 & 20.77 \\
  &   2(1)& 20.11 & 19.96 &  &   0(6)&  0.50 &  0.41\\
  &   2(5)& 36.19 & 36.59 &  &  0(10)& 37.56 & 37.96\\
  &  10(5)& 39.63 & 39.49 &  & 10(10)& 41.08 & 40.85 \\
 6&   0(0)& 22.32 & 22.13 &18&   0(0)& 19.89 & 19.87\\
  &   0(6)& 38.05 & 38.37 &  &  0(18)& 37.56 & 38.37\\
  &  12(6)& 39.63 & 39.49 &  & 36(18)& 42.06 & 41.75 \\
\hline
\hline
\end{tabular}
\end{table}

\begin{figure}[htbp]
\begin{center}
\includegraphics[height=11cm,width=6cm]{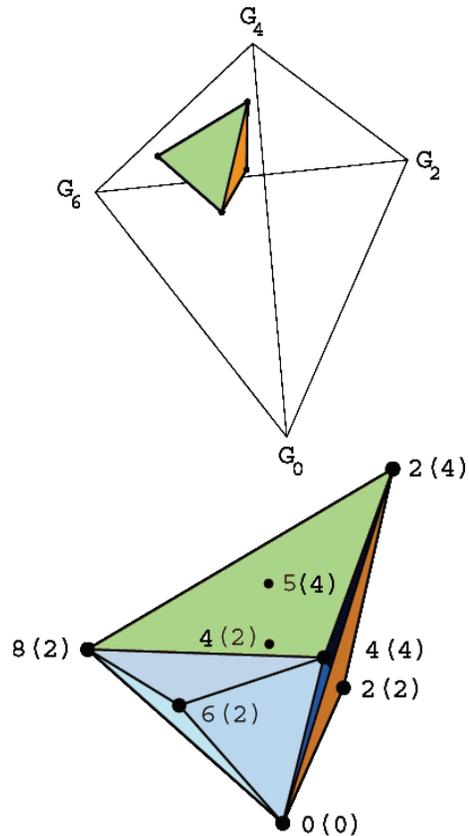}
\caption{The polyhedron for a system of $n=4$ $j={7\over2}$
 fermions. The upper part indicates the position of the
 polyhedron in the tetrahedral coordinate system.
 The left part shows the same (enlarged) polyhedron with vertices that correspond to the states $J(\nu)$ that can become the
ground state with a probability given by the exterior angle.
 Two additional states with $J(\nu)=4(2)$ and 5(4) lie inside or on the face of the polyhedron and are never the ground
state. For representation purposes the polyhedral face 0(0)--2(4)--8(2)
 has been removed in the left part.} \label{polyhedron}
\end{center}
\end{figure}

A second example concerns a system of four $j={7\over2}$ fermions
which was discussed recently by Zhao and Arima~\cite{ZHA01}.
In this case there are four interaction matrix elements,
$G_0$, $G_2$, $G_4$, $G_6$ with
$G_L\equiv\langle j^2L|\hat H_2|j^2L\rangle$,
and this leads to a problem
that can be represented in three-dimensional space.
Any state in the $j={7\over2}$ shell
can be labeled with particle number $n$, seniority $\nu$,
and total angular momentum $J$
with analytically known expansion coefficients
$b^L_{n\nu J}$~\cite{TAL93}.
For $n=4$ there are eight different states
and the corresponding coefficients $b^L_{n\nu J}$
are given in Ref.~\cite{ZHA01}.
These eight states can be represented
in three-dimensional space
spanned by the three differences
$G_0-G_6$, $G_2-G_6$, and $G_4-G_6$,
and six of them define a convex polyhedron
(see Fig.~\ref{polyhedron}).
The two remaining points
[corresponding to the state $J(\nu)=4(2)$ and 5(4)]
are inside this polyhedron and are never the ground state.
The relation between the geometry of the polyhedron
and the probability $P^{(3)}_v$ of each vertex state to be the ground state
can again be inferred from a few simple examples.
The relevant quantity in this case
is $\sum_{f\ni v}\theta_{vf}$
where the sum is over all faces that contain the vertex $v$
and $\theta_{vf}$ is the angle at vertex $v$ in face $f$.
A few examples with regular polyhedrons
(see the $m=3$ polyhedrons in Table~\ref{polytopes})
demonstrate that the relation is
\begin{equation}
P^{(3)}_v=
{1\over2}-{1\over{4\pi}}\sum_{f\ni v}\theta_{vf}.
\label{p3d}
\end{equation}
Table~\ref{j7shell} compares the probabilities
for the different states to become the ground state
as calculated in various approaches.
The second column gives the analytic results
obtained from~(\ref{p3d})
while the third column lists numerical results
obtained from 20000 runs with random matrix elements
with a Gaussian distribution.
As our code does not distinguish
between states with the same angular momentum $J$
but different seniority $\nu$,
only the summed probability for each $J$ is given.
The last column
shows the results of Zhao and Arima~\cite{ZHA01}
who calculate the probability as a multiple integral.
\begin{table}
\centering
\caption{Probabilities $P_{n\nu J}$ (in \%)
of $n=4$ $j={7\over2}$ states
obtained with the analytic formula~(\protect\ref{p3d}),
from a numerical calculation,
and from the integral representation
of Zhao and Arima~\protect\cite{ZHA01}.}
\label{j7shell}
\smallskip
\begin{tabular}{rrrr}
\hline 
\hline 
&\multicolumn{3}{c}{Probability $P_{n\nu J}$}\\
\cline{2-4}
$J(\nu)$&Analytic&Numerical&Integral\\
\hline
0(0)&    18.33& 18.38& 18.19\\
2(2)&     1.06&   ---&  0.89\\
2(4)&    33.22&   ---& 33.25\\
2(2\&4)& 34.28& 34.88& 34.14\\
4(2)&        0&   ---&  0.00\\
4(4)&    23.17&   ---& 22.96\\
4(2\&4)& 23.17& 22.83& 22.96\\
5(4)&        0&  0.00&  0.00\\
6(2)&     0.05&  0.07&  0.02\\
8(2)&    24.16& 23.83& 24.15 \\
\hline 
\hline 
\end{tabular}
\end{table}

These notions can be generalized in several ways.
The first is towards energies
that depend on a set of continuous variables $\{t_1,\dots,t_q\}$
as follows [compare with Eq.~(\ref{energ})]:
\begin{equation}
E(t_1,\dots,t_q)=
\sum_Lb_L(t_1,\dots,t_q)G_L.
\label{continuous}
\end{equation}
 The analogous problem now consists in the determination of the probability density
 $dP(t_1,\dots,t_q)$ for obtaining the lowest energy at $\{t_1,\dots,t_q\}$ with random
 interactions $G_L$. We assume by way of example that the number of variables $q$ is one less
 than the number $N$ of interactions $G_L$, $q=N-1$. In that case
 Eq.~(\ref{continuous}) represents a $q$-dimensional hypersurface $\Sigma^q$ embedded
 in a $(q+1)$-dimensional Euclidean space {\bf E}$^{q+1}$ (the metric follows from the
 covariance matrix $\langle G_L G_{L'} \rangle=\delta _{LL'}$).
 Let us suppose that $\Sigma^q$ is the closed orientable manifold.
 It can then be shown that the probability density is given by Gauss' spherical map~\cite{KoNo69} $\Sigma^q \to S^q$ where $S^q$ is
 a $q$-dimensional hypersphere of a unit radius.
If the degree of the spherical map is one (as it is for closed convex surfaces), the probability density reads
\begin{equation}
dP(t_1,\dots,t_q)=
{\frac{1}{S_q}}
K_q(t_1,\dots,t_q)dv,
\label{dP}
\end{equation}
where $K_q$ is the {\em Gaussian curvature} of $\Sigma^q$,
$S_q$ is the volume of the unit hypersphere $S^q$
and $dv$ denotes an infinitesimal element of $\Sigma^q$.
In the simplest example of one parameter $t_1\equiv t$
and two interactions $G_1$ and $G_2$,
the energy is parametrized as a curve in a plane.
It can then be shown that the probability density
is given by $dP(t)=(2\pi)^{-1}K_1(t)ds$,
where $ds$ is the infinitesimal arc length.
In fact, this formula is also valid for piecewise curves
and precisely leads to the result~(\ref{p2d}) for a polygon.
 The validity of the result can be checked by comparing the probability
 obtained by integration of (\ref{dP}) over a part of $\Sigma^q$
 to the numerically calculated one.
As an example we discuss a two-dimensional ellipsoid in {\bf E}$^3$
with one semi-axis $c$ different from the others $a$.
The probability associated with the part of the surface
with spherical coordinates $(\theta,\phi)$
satisfying $0\leq\theta\leq\theta_0$
and $0\leq\phi\leq\phi_0$ is given by
\begin{eqnarray}
P(\theta_0,\phi_0)
=\frac{\theta_0}{4\pi}
\left(1-
\frac{a\cos\phi_0}{\sqrt{c^2+(a^2-c^2)(\cos{\phi_0})^2}}\right).
\label{3d}
\end{eqnarray}
This expression has been compared with the numerically calculated probability; the difference is close to zero. 
We have analyzed likewise the case
of a three-dimensional hyperellipsoid in {\bf E}$^4$,
showing that our approach
can be generalized to higher dimensions.
These results also open up the possibility
 for an extension to higher-dimensional polytopes,
 by replacing the right-hand side of (Eq.~{\ref{dP})
 with an appropriate characteristic for each vertex of the polytope.
 Indeed, it can be shown that the probabilities $P^{(m)}_v$ in~(\ref{p2d}) and~(\ref{p3d})
 are related to the {\em exterior angle} at vertex $v$~\cite{GRU67,BAN67}
 of either a convex polygon ($m=2$) or a convex polyhedron ($m=3$).

We believe that,
although derived for a restricted form of interaction Hamiltonians,
these results suggest that generic $n$-body quantum systems,
interacting through two-body forces,
can be associated with a geometrical shape
defined in terms of CFPs or generalized coupling coefficients.
Geometry thus arises as a consequence of strong correlations
implicit in such systems
and is independent of particular two-body interactions.
Random tests can be understood in this context
as sampling experiments on this geometry.
In this Letter we have shown
that geometric aspects of $n$-body systems
determine some of their essential characteristics.
In particular, for diagonal Hamiltonians
surface curvature defines probability to be the ground state.
Other correlations could also be related to geometric features.
These results generalize and put onto a firm basis
the previous work which hinted at a purely geometric interpretation
of randomly interacting boson systems~\cite{BIJ00,KUS00},
as well as provide an explanation for the method of Zhao and collaborators~\cite{ZHA01}.
In fact, in the latter reference, the authors have advanced some
qualitative reasons for certain states to dominate and later provided an
approximate procedure (which is not always accurate~\cite{ZHA02}) to estimate the
ground state probabilities, although no reason was offered for its success.
Our study, at least for the case where the Hamiltonian is diagonal, clearly
shows that the procedure of Zhao et al is equivalent to a projection of the considered above
polyhedra on the axes defined by the two-body matrix elements, 
which tend to correlate well with the angles we introduce. 
This connection will be elaborated in detail elsewhere~\cite{CHAU}.
Further work is required to fully explore the geometry
and its consequences for our understanding of $n$-body dynamics.

Acknowledgments: AF is supported by CONACyT, Mexico,
under project 32397-E.
NAS thanks L.~Vanhecke for a helpful discussion.

\end{document}